# Low Threshold Parametric Decay Back Scattering Instability in Tokamak ECRH Experiments

E.Z. Gusakov and A.Yu. Popov

*Ioffe Institute, St.-Petersburg, Russia*

Abstract. The experimental conditions leading to substantial reduction of backscattering decay instability threshold in ECRH experiments in toroidal devices are analyzed. It is shown that drastic decrease of threshold is provided by non monotonic behavior of plasma density in the vicinity of magnetic island and poloidal magnetic field inhomogeneity making possible localization of ion Bernstein decay waves. The corresponding ion Bernstein wave gain and the parametric decay instability pump power threshold is calculated.

**Introduction**

Electron cyclotron resonance heating (ECRH) at power level of up to 1 MW in a single microwave beam is often used in present day tokamak and sellarator experiments and planed for application in ITER for neoclassical tearing mode control. Parametric decay instabilities (PDI) leading to anomalous reflection or absorption of microwave power are believed to be deeply suppressed in tokamak MW power level ECR O-mode and second harmonic X-mode heating experiments utilizing gyrotrons [1]. Therefore it is taken for granted that wave propagation and absorption in these experiments is well described by linear theory and thus predictable in detail.

According to theoretical analysis of PDI thresholds [1-3], the typical RF power at which these nonlinear effects can be excited at tokamak plasma parameters is around 1 GW, which is only possible with a free electron laser microwave source. The physical reason for such a deep suppression is provided by strong convective losses of daughter waves from the decay region either in the plasma inhomogeniety direction or along the magnetic field. In the first case the daughter waves amplification in the narrow region, where the decay condition $\Delta k = k_1 - k_2 - k_0 = 0$, ($k_1$, $k_2$ and $k_0$ - wave numbers of two daughter waves and pump, correspondingly), is fulfilled in inhomogeneous plasma, is described by the so called Piliya – Rosenbluth coefficient [4-6]

$$S_{PR} = \exp\left\{\frac{\pi \gamma_0^2 \ell^2}{v_1 v_2}\right\} \quad (1)$$

where $\gamma_0$ - is the maximal PDI growth rate in homogeneous plasma, proportional to the pump wave amplitude, $v_1$ and $v_2$ - the daughter wave group velocities and $\ell^2 = \left|d(\Delta k)/dx\right|^{-1/2}$. As it is clear from this formula the PDI threshold can be lowered by the growth of the pump field or/and by decrease of the daughter wave group velocity. Both effects occur in the case of induced backscattering in the vicinity of the pump wave upper hybrid resonance [7] explaining easy PDI excitation in EBW heating experiments. Until recently it was the only situation where backscattering PDI was observed at 100 kW RF power level [8]. However last year the first observations of the backscattering signal in the 200 – 600 kW level second harmonic ECRH experiment at TEXTOR tokamak were reported [9,10]. This signal down shifted in frequency by approximately 1 GHz, which is close to the lower hybrid frequency under the TEXTOR conditions, was strongly modulated in amplitude at the *m=2* magnetic island frequency. This observation performed at the modest RF power under conditions when no UHR was possible for the pump wave provides an indication that probably a novel low threshold mechanism of the PDI excitation is associated with the presence of a magnetic island.

In the present paper the experimental conditions not related to the UHR leading to substantial reduction of backscattering decay instability threshold in tokamak ECRH experiments are analyzed. The parametric decay of the pump X-mode into backscattering X-mode and ion Bernstein IB wave is considered accounting for non monotonous density profile in the vicinity of magnetic island O-point and poloidal magnetic field inhomogeneity. The IB wave amplification coefficient and the PDI threshold are obtained. The later is shown to be substantially (4 orders of magnitude) smaller than that provided by the standard theory [1-3].

**The basic equations**

To elucidate the physics of the PDI threshold lowering we shall analyze the most simple (but nevertheless relevant to the experiment [9]) three wave interaction model in which both the X-mode pump and high frequency X-mode decay wave propagate almost perpendicular to the magnetic field in the density inhomogeneity direction (*x*). Moreover, for the sake of simplicity we assume the pump frequency exceeding both the electron cyclotron and plasma frequency, which is the case in the TEXTOR experiments. We neglect also a weak dependence of the high frequency wave numbers $k_{ix}$ and $k_{sx}$ on coordinate that allows us to introduce the pump wave in the form $E_{iy} = a_i \cdot \exp\left(ik_{ix}x - i\omega_i t - (y^2 + z^2)/(2\rho^2)\right)$ describing a wide microwave beam propagating from the launching antenna inwards plasma with an amplitude $a_i = \sqrt{8\pi P_i/(\pi\rho^2 c)} = const$, where $P_i$ is the pump wave power and $\rho$ is the beam radii and axes *y* and *z* correspond to poloidal and toroidal directions. The equation describing the backscattered wave generation and its convective losses from the decay region in the density inhomogeneity direction takes a form

$$\left(\frac{\partial^2}{\partial x^2} + k_{sx}^2\right)E_{sy} = i\frac{4\pi\omega_s}{c^2}j_{sy} \qquad (2)$$

where the nonlinear current in (2) is given by a product of a electron density perturbation $\delta n_\Omega$ produced by a low-frequency ($\Omega = \omega_i - \omega_s \ll \omega_i$) small scale decay wave and the quiver electron velocity $u_{iy}$ in the pump wave field $j_{sy} = e\delta n_\Omega u_{iy}$. Assuming electrostatic low frequency wave $\vec{E} = -\vec{\nabla}\varphi \exp(i\Omega t)$ we get the nonlinear current density perturbation in the form

$$j_{sy} \simeq -\frac{e}{m_e c^2} \frac{\omega_{pe}^2}{\omega_{ce}^2} \left(\frac{\partial^2}{\partial x^2}\varphi\right) E_{iy} \tag{3}$$

We seek a solution of (2) in the form $E_{s,y} = a_s(\vec{r})\exp(-ik_{s,x}x - i\omega_s t)$, where an amplitude $a_s(\vec{r})$ varies slowly due to non-linear interaction in the decay layer. The low frequency daughter wave potential is described by the integral equation

$$\int_{-\infty}^{\infty} d\vec{r}' D(\vec{r} - \vec{r}', (\vec{r} + \vec{r}')/2) \varphi(\vec{r}') = 4\pi \rho_\Omega \tag{4}$$

|In weakly inhomogeneous plasma the kernel of this equation, exhibiting much stronger dependence on the first argument $\vec{r} - \vec{r}'$ than on second - $(\vec{r} + \vec{r}')/2$, associated with the plasma inhomogeneity, can be represented in terms of homogeneous plasma theory as

$$D(\vec{r} - \vec{r}', (\vec{r} + \vec{r}')/2) = (2\pi)^{-3} \int_{-\infty}^{\infty} D(\vec{q}, (\vec{r} + \vec{r}')/2) \exp[i\vec{q}(\vec{r} - \vec{r}')] d\vec{q}, \text{ where}$$

$$D(\vec{q}, (\vec{r} + \vec{r}')/2) = q^2 \left(1 + \chi_e(\vec{q}, (\vec{r} + \vec{r}')/2) + \chi_i(\vec{q}, (\vec{r} + \vec{r}')/2)\right) = D_\perp + D_\parallel \tag{5}$$

The electron susceptibility $\chi_e$, entering this expression is provided by cold homogeneous plasma model in the form $\chi_e = q_\perp^2/q^2 \cdot \omega_{pe}^2/\omega_{ce}^2 - q_\parallel^2/q^2 \cdot \omega_{pe}^2/\Omega^2$, where wave vector components along and across magnetic field are given correspondingly by $q_\parallel^2 = (q_z \cos\varphi + q_y \sin\varphi)^2$ and $q_\perp^2 = q_x^2 + (q_y \cos\varphi - q_z \sin\varphi)^2$. For the ion susceptibility $\chi_i$ we use representation derived in [11]

$$\chi_i = \frac{2\omega_{pi}^2}{q^2 v_{ti}^2} \left[1 - X\left(\frac{\Omega}{q_\perp v_{ti}}\right) - \left(\cot\left(\frac{\pi\Omega}{\omega_{ci}}\right) + \frac{i}{\sqrt{\pi}} \frac{\omega_{ci}}{|q_\parallel|v_{ti}} \sum_{m=-\infty}^{\infty} \exp\left(-\frac{(\Omega - m\omega_{ci})^2}{q_\parallel^2 v_{ti}^2}\right)\right) Y\left(\frac{\Omega}{q_\perp v_{ti}}\right)\right] \tag{6}$$

where $X(\xi) + iY(\xi) = \frac{\xi}{\sqrt{\pi}} \int_{-\infty}^{\infty} \frac{\exp(-t^2)}{t - \xi - io} dt$. Below we also use the following notation $D_\perp = q_\perp^2 \omega_{pe}^2/\omega_{ce}^2 + q^2 \chi_i$ and $D_\parallel = -q_\parallel^2 \omega_{pe}^2/\Omega^2$. The nonlinear charge density $\rho_\Omega$ in (4) responsible for coupling of low and high frequency waves is provided by a ponderomotive force which, in the LH frequency range takes a form

$$\rho_\Omega = \frac{1}{4\pi} \frac{e}{m_e} \frac{\omega_{pe}^2}{\omega_{ce}^2 \omega_i^2} \frac{\partial}{\partial x}\left[E_{iy}^* \frac{\partial E_{sy}}{\partial x} + E_{sy} \frac{\partial E_{iy}^*}{\partial x}\right] \tag{7}$$

Taking into account that, according to expression (1), the PDI amplification is enhanced when the daughter wave group velocity decreases we shall consider solutions of (4) in the vicinity of the low frequency wave turning point in x direction ($x = x_0$). Here conditions $D|_{q_{x0}, \Omega_0, x_0} \equiv D|_0 = 0$, $\partial D/\partial q_x|_0 = 0$ hold for wave at frequency $\Omega = \Omega_0$, wave number $q_x = q_{x0}$ and its group velocity goes to zero. As it is seen in Figure 1, due to periodicity of $\cot(\pi\Omega/\omega_{ci})$ function solution of equations $D\big|_0 = 0$ and

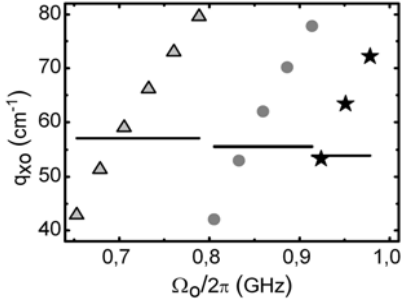

FIG.1 (color online). IB wave turning points at $T_i = 600$ eV; $n = 1 \times 10^{13}$ cm$^{-3}$ (triangles); $n = 2 \times 10^{13}$ cm$^{-3}$ (circles), $n = 3 \times 10^{13}$ cm$^{-3}$ (stars); solid lines - $k_{i,x}(x_0) + k_{s,x}(x_0)$ at the corresponding density.

$\partial D/\partial q_x\big|_0 = 0$ is not unique and it is easy to find one approximately satisfying the pump Bragg backscattering condition ($q_{x0} \simeq k_{i,x} + k_{s,x}$). Moreover, in attempt to explain the extremely low pump power level at which backscattering correlated to magnetic activity was observed in [9, 10] we shall make use of non monotonous profile of plasma density often observed by diagnostics possessing high temporal and spatial resolution in the vicinity of magnetic island. (Such an observation was made, in particular, in some experiments at TEXTOR [12, 13]. Namely, we assume, as it is shown in experiment [13], that a local density maximum is associated with the O-point and therefore conditions $\partial D/\partial x\big|_0 = 0$ and $q_{x0}^2 L_{nx}^{-2} \equiv \partial^2 D/\partial x^2\big|_0 > 0$ hold there. In these circumstances two nearby turning points ("warm" to "hot" mode) exist in plasma for high harmonic IB wave and it can be trapped in $x$ direction if additional condition $\partial^2 D/\partial q_x^2\big|_0 < 0$ holds. (If the opposite condition $\partial^2 D/\partial q_x^2\big|_0 > 0$ holds in the island, IB wave trapping can occur in the local density minimum also accompanying magnetic island, as it is shown in Fig. 2.) The corresponding density profile and IB wave dispersion curve $q_x(x)$ calculated for frequency $\Omega_0 = 0.85$ GHz, ion temperature $T_i = 600$ eV is shown in Fig.2. The described possibility of IB wave localization in a plasma waveguide provides an argument in favor of the PDI threshold lowering in the case of the waveguide eigen mode excitation. The physical reason for it is related to the suppression of convective wave energy losses in the inhomogeniety direction. It is important to note that due to magnetic field dependence on major radius IB wave trapping is possible also in the poloidal direction. It occurs when IB wave is excited with small parallel wave number close to the equatorial plane at the low magnetic field side of the torus.

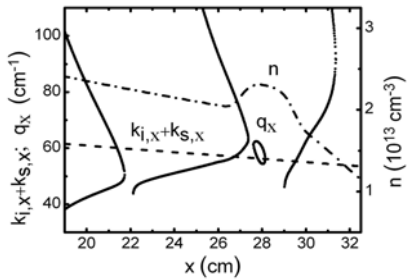

FIG.2 (color online). The IB wave dispersion curves (solid line) and plasma density (dash dotted line) in the magnetic island vicinity. Dashed line – $k_{i,x} + k_{s,x}$.

In the vicinity of magnetic island O-point, situated in the equatorial plane of the torus and coincident to the IB wave turning point, equation (4) can be reduced to differential equation

$$\left\{\frac{\partial D}{\partial \Omega}\bigg|_0 \delta\Omega - \frac{\partial^2 D}{2\partial q_x^2}\bigg|_0 \frac{\partial^2}{\partial x^2} + \sin^2\phi \frac{\omega_{pe}^2}{\Omega^2}\bigg|_0 \frac{\partial^2}{\partial y^2} + q_{x0}^2 \frac{(x-x_0)^2}{L_{nx}^2} - q_{x0}^2 \frac{y^2}{L_b^2} - \delta D\right\} b = 4\pi\rho_\Omega \exp\left[iq_{x0}x - iq_z(y\cot\phi - z)\right] \quad (8)$$

where a substitution $\varphi = b(x,y)\exp[-iq_{x0}x + iq_z(y\cdot\cot\phi - z)]$ assuming wave propagation almost perpendicular to magnetic field was made and the following notations were introduced:

$$\tan\phi = B_y/B_z\,;\; L_b^{-2} = \left.\frac{\omega_{pi}^2(x)}{q_x^2 v_{ti}^2}\csc^2\left(\pi\frac{\Omega}{\omega_{ci}}\right)Y\left(\frac{\Omega}{q_x v_{ti}}\right)\frac{\pi\Omega}{\omega_{ci}}\frac{1}{rR}\right|_0\,;\; r \text{ and } R \text{ are minor and major radii;}$$

$$\delta D = \frac{q_z^2}{4\sin^2\phi}\left[\left.\left(\frac{1}{q_x}\frac{\partial^3 D}{\partial q_x^3} - \frac{2}{q_x^2}\frac{\partial^2 D}{\partial q_x^2}\right)\right|_0 \frac{\partial^2}{\partial x^2} + \left.\frac{3}{q_x^2}\frac{\partial^2 D}{\partial q_x^2}\right|_0 \frac{\partial^2}{\partial y^2}\right] + \frac{2i\omega_{pi}^2}{\sqrt{\pi}v_{ti}^2}\frac{\omega_{ci}}{|q_\parallel|v_{ti}}\sum_{m=-\infty}^{\infty}\exp\left(-\frac{(\Omega - m\omega_{ci})^2}{q_\parallel^2 v_{ti}^2}\right)Y\left(\frac{\Omega}{q_\perp v_{ti}}\right)$$

is the perturbation describing convective losses in toroidal direction and IBW damping, $\delta\Omega = \Omega - \Omega_0$. In

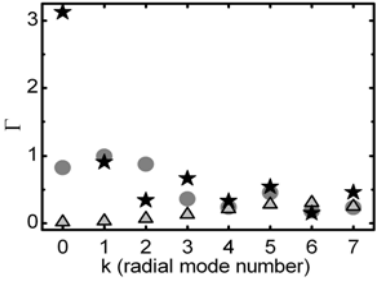

FIG.3 (color online). Dependence of the IB wave gain on the radial mode number; $P = 400$ kW, $T_i = 600$ eV; Triangles - $n = 1\times10^{13}\,\text{cm}^{-3}$, $\Omega_0/2\pi = 0.68$ GHz; Circles - $n = 2\times10^{13}\,\text{cm}^{-3}$, $\Omega_0/2\pi = 0.83$ GHz; Stars - $n = 3\times10^{13}\,\text{cm}^{-3}$, $\Omega_0/2\pi = 0.92$ GHz.

this consideration we neglected weak poloidal and moreover very weak toroidal density inhomogeniety in the magnetic island and therefore suppose wave number $q_z$ constant.

**The PDI analyses and discussion**

Assuming the IB wave damping, PDI pumping and convection in toroidal direction weak we account for them using the perturbation theory approach [14]. In the zero order approximation we neglect $\delta D$ and $\rho_\Omega$ in (8) and obtain equation which can be solved by separation of variables. The corresponding expression for the IB eigen mode trapped in radial and poloidal direction and possessing $q_z = 0$, which corresponds to suppressed convective losses in toroidal direction is

$$b(x,y) = \varphi_k(x)\varphi_l(y) = H_k\left(\frac{x-x_0}{\delta_x}\right)H_l\left(\frac{y}{\delta_y}\right)\exp\left[-\left(\frac{x-x_0}{\sqrt{2}\delta_x}\right)^2 - \left(\frac{y}{\sqrt{2}\delta_y}\right)^2\right] \quad (9)$$

where $H_k$ is standing for Hermitian polynomial, the size of the IB mode localization region is $\delta_x = (L_{nx}/q_{x0})^{1/2}\left(\partial^2 D/\partial q_x^2\big|_0/2\right)^{1/4}$, $\delta_y = \left.\left(\sin\phi(L_b/q_{x0})^{1/2}\omega_{pe}/\Omega\right)\right|_0^{1/2}$ and the exact value of the mode frequency is determined by quantization condition

$\delta\Omega_{k,l} = \partial D/\partial\Omega\big|_0^{-1}\left(\sqrt{\partial^2 D/\partial q_x^2\big|_0/2}\cdot q_{x0}/L_{nx}\cdot(2k+1) - \sin\phi\sqrt{\omega_{pe}^2/\Omega^2}\big|_0\cdot q_{x0}/L_b\cdot(2l+1)\right)$. At the next step of the perturbation analysis procedure we account for IB wave damping, PDI pumping and convection in toroidal direction. Expressing the BS wave amplitude from (2) in terms of the IB wave potential and introducing $\Delta K = q_{x0} - k_{i,x} - k_{s,x}$ we obtain the nonlinear charge density in the form

$$\rho_\Omega[b(x,y)]\exp(iq_{x0}x - iq_z(y\cot\phi - z)) = -i\frac{q_{x0}^3}{16\pi}\frac{\omega_{pe}^4}{\omega_i^2\omega_{ce}^2}\frac{|a_i|^2}{H^2}\exp\left(-\frac{y^2+z^2}{\rho^2}\right)\int_{-\infty}^{x}dx'\exp(i\Delta K(x-x'))b(x',y)\cdot$$

Substituting the zero order solution (9) into this expression and in $\delta D$ and requiring no variation of the eigen frequency $\delta\Omega_{kl}$ with the accuracy up to the first order we obtain equation for the toroidal wave number determining damping or PDI amplification of the IB wave

$$\int_{-\infty}^{\infty}\int_{-\infty}^{\infty}\varphi_k(x)\varphi_l(y)\{\delta D[\varphi_k(x)\varphi_l(y)]-4\pi\rho_\Omega[\varphi_k(x)\varphi_l(y)]\exp(iq_{x0}x-iq_z(y\cot\phi-z))\}dydx=0 \quad (10)$$

The necessary condition for the PDI onset is provided by excess of the pump contribution to $q_z$ over the damping contribution. When the ion cyclotron harmonic is far from the IB wave trapping region ($(\Omega-m\omega_{ci})/(q_z v_{ti})\gg 1$) the damping is negligible and the imaginary part of toroidal wave number $q_z''$ is given by expression

$$q_z''=|\sin\phi|q_{x0}^{5/2}\delta_x^{3/2}\frac{\sqrt{2}}{2}\frac{\omega_{pe}^2}{\omega_i\omega_{ce}}\frac{\Omega_0}{\omega_{pi}}\frac{a_i}{H}\alpha_{k,l}, \quad (11)$$

where

$$\alpha_{k,l}^2=\frac{(k!2^k l!2^l)^{-1}\varphi_k^2(\Delta K\delta_x)\int_{-\infty}^{\infty}dy[\exp(-y^2/\rho^2)\varphi_l^2(y)]}{\left(k+\frac{1}{2}\right)\left(q_x\frac{\partial^3 D}{\partial q_x^3}-2\frac{\partial^2 D}{\partial q_x^2}\right)\Big|_0+\left(l+\frac{1}{2}\right)\frac{\delta_x^2}{\delta_y^2}3\frac{\partial^2 D}{\partial q_x^2}\Big|_0}, \quad \alpha_{0,0}^2\simeq\frac{2\sqrt{\pi}\exp[-(\delta_x\Delta K)^2]}{\left(q_x\cdot\frac{\partial^3 D}{\partial q_x^3}-\left(2-3\cdot\frac{\delta_x^2}{\delta_y^2}\right)\frac{\partial^2 D}{\partial q_x^2}\right)\Big|_0\sqrt{1+\frac{\delta_y^2}{\rho^2}}}.$$

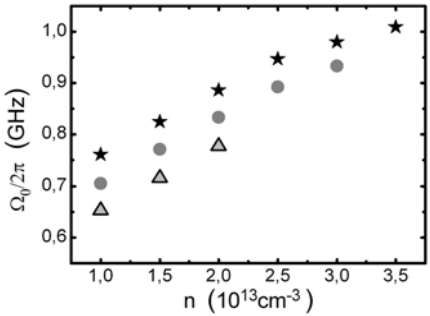

FIG.4 (color online). Dependence of frequency corresponding to maximal IB wave gain on plasma density. (Triangles - $T_i=300$ eV; Circles - $T_i=600$ eV; Stars - $T_i=900$ eV)

In the case of wide pump beam $\rho>\sqrt{2l+1}\delta_y\cot\phi$ the sufficient condition for the PDI onset is given by expression $\Gamma=2\sqrt{2\pi}\rho q_z''>1$ determining large enough gain of the IB wave when propagating across pump beam along toroidal direction. In the opposite case at $\rho<\sqrt{2l+1}\delta_y\cot\phi$ the poloidal gain over the IB mode localization region provided by exponential factor $\exp[iq_z y\cdot\cot\phi]$ dominates and the PDI threshold takes a form $\Gamma=2\sqrt{2l+1}\delta_y\cot\phi\cdot q_z''>1$. Dependence of this gain on the IB radial mode number, computed accounting for the IB wave damping, is shown in Fig.3 for $l=0$, TEXTOR parameters and different plasma densities. As it is seen, because of decay condition mismatch $\Delta K\neq 0$ the gain is not always maximal for the fundamental mode. As it is shown in Fig.4, the IB frequency corresponding to the maximal gain is growing with the plasma density. This dependence is consistent with corresponding dependence of the back scattered wave frequency shift observed in [10].

The PDI power threshold provided by condition $\Gamma>1$ is given by formula

$$P_{th}=\frac{cH^2}{2\pi\alpha_{k,l}^2}\frac{\omega_i^2\omega_{ce}^2}{\omega_{pe}^4}\frac{\omega_{pi}^2}{\Omega_0^2}\left(\frac{\rho^2}{(2l+1)\delta_y^2}\frac{1}{q_x^5\delta_x^3}\right)\Big|_0 \quad (12)$$

In the case of typical TEXTOR experimental parameters ($H=19$ kGs, $f_i=140$ GHz, $n=3\times 10^{13}$ cm$^{-3}$, $T_i=600$ eV, $\rho=1$ cm) assuming for magnetic island density variation $\delta n/n=0.1$ and width $w=3$ cm, as measured in [5], we obtain for the fundamental IB mode (*k=0, l=0*) the threshold value $P_{th} \approx 45$ kW which is overcome in the experiment. The corresponding frequency shift is $\Omega_0 = 0.92$ GHz, $\delta\Omega_{0,0}/(2\pi) = 10$ MHz, $\delta_y = 0.8$ cm, $\delta_x = 0.6$ cm and $\Delta K \delta_x \ll 1$.

**Conclusions**

The obtained drastic, compared to predictions of the standard theory [1-3], decrease of the PDI threshold is explained by complete suppression of IB wave radial and poloidal convective losses and their substantial reduction in the third direction. This mechanism is based first of all on magnetic island confinement properties which, we believe, are not specific for the TEXTOR experimental conditions and may lead to easy PDI excitation in ECRH experiments in other toroidal devices where magnetic islands usually exist. Moreover it should be mentioned that not only magnetic island, but also drift wave density perturbations, which are as well elongated along magnetic field, in the case of intensive enough turbulence can result in IB wave trapping. Similar effect leading to reduction of PDI threshold can occur also on the plasma discharge axis. Backscattering PDI can lead to reduction of ECRH efficiency and change of its localization. It should be also underlined that absorption of parametrically driven IB wave can be responsible for ion acceleration often observed in ECRH experiments (see [15] and references there).

**Acknowledgments**

Financial support by RFBR grants 10-02-90003-Bel, 10-02-00887, NWO-RFBR Centre of Excellence on Fusion Physics and Technology (grant 047.018.002) and scientific school grant- 6214.2010.2 is acknowledged.